\documentclass[twocolumn,superscriptaddress,aps,prb,showpacs,showkeys,amsmath]{revtex4}

\usepackage{graphicx}												
\usepackage[thinspace,thinqspace]{SIunits}	
\usepackage[T1]{fontenc}
\usepackage[latin1]{inputenc}
\usepackage{booktabs}
\usepackage{multirow}
\usepackage{array}

\newcommand{\dd}{\mathrm{d}} 															
\newcommand{\fracpd}[2]{\frac{\partial #1}{\partial #2}} 	
\newcommand{\HE}{Hall effect}															
\newcommand{\HC}{Hall coefficient}												
\newcommand{\YRS}{YbRh$_2$Si$_2$}													
\newcommand{\YIS}{YbIr$_2$Si$_2$}													
\newcommand{\LRS}{LuRh$_2$Si$_2$}													
\newcommand{\EF}{\ensuremath{E_{\text F}}}								
\newcommand{\RH}{\ensuremath{R_{\text H}}}								
\newcommand{\RT}{\ensuremath{\RH (T)}}										
\newcommand{\RHO}{\ensuremath{\RH (T\rightarrow 0)}}			
\newcommand{\rhoH}{\ensuremath{\rho_{\text H}(B)}}				
\newcommand{\HEunit}[1]{\unit{#1 \times 10^{-10}}\cubicmetre\per\coulomb}
\newcommand{\rydberg}{\ensuremath{Ry}}
\newcommand{\percent}{\ensuremath{\%}}
\newcommand{\SymbolHeight}{0.75em}
\newcommand{\mK}{\milli\kelvin}
\newcommand{\TN}{\ensuremath{T_{\text N}}}
\newcommand{\TRH}{\ensuremath{\tilde R_{\text H}}}				

\newcolumntype{V}{>{\centering\arraybackslash} m{.4\linewidth} }

\begin{document}

\preprint{Manuscript: version 2.4 Date: \today\  \thistime}

\title{Hall effect of YbRh$_2$Si$_2$ and relatives in the light of electronic structure calculations}

\author{Sven Friedemann\footnote{Present address: Cavendish Laboratory, University of Cambridge, JJ Thomson Avenue, Cambridge CB3 0HE, UK.}}
	\affiliation{Max Planck Institute for Chemical Physics of Solids, Nöthnitzer Strasse 40, 01187 Dresden, Germany}%
	\email{Sven.Friedemann@cpfs.mpg.de}
	
\author{Steffen Wirth}

\author{Niels Oeschler}
		
\author{Cornelius Krellner}
	
\author{Christoph Geibel}
	
\author{Frank Steglich}

\affiliation{Max Planck Institute for Chemical Physics of Solids, Nöthnitzer Strasse 40, 01187 Dresden, Germany}%

\author{Sam MaQuilon}
	
\author{Zachary Fisk}

\affiliation{Department of Physics and Astronomy, University of California, Irvine, CA 92697-4575, USA}

\author{Silke Paschen}

\affiliation{Institute of Solid State Physics, TU Vienna, Wiedner Hauptstraße 8-10, 1040 Vienna, Austria}

\author{Gertrud Zwicknagl}

\affiliation{Institut für Mathematische Physik, TU Braunschweig, Mendelssohnstraße 3, 38106 Braunschweig, Germany}

\date{\today }

\begin{abstract}
We report experimental and theoretical investigations of 
the Hall effect in YbRh$_2$Si$_2$ and its reference compounds LuRh$_2$Si$_2$ and YbIr$_2$Si$_2$. Based on band-structure calculations we identify two bands dominating the Hall coefficient in all these compounds. For the case of LuRh$_2$Si$_2$---the non-magnetic reference compound of YbRh$_2$Si$_2$---the temperature dependence of the Hall coefficient is described quantitatively to arise from two hole-like bands. For YbIr$_2$Si$_2$ and YbRh$_2$Si$_2$, renormalized band calculations yield two bands of opposite character. In YbRh$_2$Si$_2$ these two bands almost compensate each other. We present strong indications that sample dependences of the low-temperature Hall coefficient observed for YbRh$_2$Si$_2$ arise from slight variations of the relative scattering rates of the two bands. Minute changes of the composition appear to be the origin.
{The results of our band structure calculations reveal that a transition of the 4$f$ electrons from localized to itinerant leads to a decrease of the Hall coefficient.}
\end{abstract}

\maketitle
%
%
\section{Introduction}
The heavy-fermion compound \YRS\ has emerged as a prototypical system for the investigation of quantum critical phenomena.\cite{Gegenwart2008}  
Pronounced non-Fermi-liquid properties arise due to the proximity to a quantum critical point (QCP).\cite{Custers2003}
In its ground state, \YRS\ orders antiferromagnetically below the Néel temperature, $\TN=\unit{70}\mK$.\cite{Trovarelli2000b} By applying a small magnetic field of $B_{\text c}= \unit{60}\milli\tesla$ within the basal plane, the magnetic order is suppressed to zero temperature, thus accessing the field-induced QCP.\cite{Gegenwart2002}

Hall-effect measurements turned out to be of central importance to understand the nature of the QCP as they allow to discriminate two different theoretical scenarios\cite{Coleman2001} as discussed below.
The Hall coefficient \RH\ of \YRS\ was measured as the compound was driven from the magnetically ordered state across the QCP towards the LFL regime by increasing the magnetic field.\cite{Paschen2004,Friedemann2009a} Since anomalous contributions\cite{Fert1987} are negligible\cite{Paschen2005} at low temperatures, \RH\ is directly related to 
the Fermi surface volume. The Hall coefficient was found to exhibit a crossover linked to the QCP which resides on top of a smooth background. Since this crossover sharpens to a discontinuous jump in the extrapolation to zero temperature these results imply an abrupt change of the Fermi surface at the QCP. Such a Fermi surface reconstruction is at variance with the predictions of the standard spin-density-wave theory.\cite{Hertz1976,Millis1993,Moriya1995} Rather, the results suggest a new class of theoretical descriptions to be applied in \YRS, namely the Kondo-breakdown scenario in which the 4$f$ electrons are itinerant on the high-field side of the QCP only.\cite{Coleman2001,Si2001,Senthil2004,Pepin2007} Consequently, the Hall effect represents a key experiment to identify the unconventional nature of the quantum criticality in \YRS.


On the other hand, it was pointed out that the \HC\ of \YRS\ is not simply proportional to the inverse charge carrier concentration since the assumption of a spherical Fermi surface with a single band at the Fermi energy \EF\ is not valid in this material as shown by various band-structure calculations\cite{Wigger2007,Norman2005} and photoemission studies.\cite{Wigger2007,Danzenbacher2007} Several calculations yield multiple bands crossing \EF\ with canceling positive and negative contributions to the \HC .\cite{Jeong2006,Norman2005} Thus, the \emph{calculated} \HC\ critically depends on the method used, and it remains an outstanding challenge to interpret the measured \HC\ quantitatively in terms of band-structure calculations.

The Hall-effect measurements\cite{Friedemann2008} performed in a low-temperature setup with improved resolution on a number of \YRS\ single crystals reproduced the results on a crystal used in Ref.~\onlinecite{Paschen2004}. Other crystals of different quality, however, show strong sample dependences below \unit{20}\kelvin. Although the critical crossover is found to be virtually independent of sample quality, it remains to be understood why the underlying background exhibits such strong sample dependences.\cite{Friedemann2009a}

Here, we present experimental and theoretical progress which helps to refine our understanding of the \HE\ in \YRS. The issue of the low-temperature sample dependences as well as the characteristics of \RT\ in \YRS\ are addressed. Comparison with the Hall-effect data of both the non-magnetic reference compound \LRS\ and the heavy-fermion compound \YIS\ allows us to discriminate various contributions to the \HC . Our renormalized band-structure calculations yield excellent agreement with the experimentally determined \HC\ for \YIS\ and \LRS. They provide a reliable basis to understand the sample dependences in \YRS . Moreover, enable us to relate the field induced crossover of the Hall coefficient to a change in the carrier concentration \cite{Paschen2004,Friedemann2009a}. 

We present the details and the results of the electronic structure calculations in section \ref{sec:BSC}. This includes the calculation of the Hall coefficient in section \ref{subsec:Calc_HC}. In section \ref{sec:HE} the results of the electronic transport measurements on \LRS, \YIS, and  \YRS\ are presented and discussed in the light of determined electronic band structures.

%
%
\section{band-structure calculations}
\label{sec:BSC}
%
%
\subsection{Models of electronic structure}
%
%
\subsubsection{Local moment regime}
\label{subsubsec:local_moment_regime}
We begin by calculating the electronic structure of \YRS\ and \YIS, assuming that the
Yb ions are in the 4$f$$^{13}$ configuration. With this approximation we model the Fermi surface and the quasiparticle bands in the local moment regime.
As there are exactly 13 4$f$ electrons or one 4$f$ hole per Yb site the single-particle excitations of the 4$f$ shell involve valence transitions $4f^{13}\to 4f^{14}$ and
$4f^{13}\to 4f^{12}$ which occur at high energies only.
Consequently, the 4$f$ degrees of freedom do not contribute to the
low-energy excitations in the vicinity of the Fermi surface. In this
energy range, the single-particle excitations are derived from the  weakly correlated
(non-$f$) conduction states which form coherent Bloch states. We determine
the dispersion of these bands by standard band-structure calculations.
The effective potentials are generated self-consistently within the
Local Density Approximation (LDA) to density functional theory.
The strong Coulomb repulsion among the 4$f$ electrons which suppresses
charge fluctuations is implicitly accounted for by treating the 4$f$
electrons as part of the ion core assuming that they do not hybridize
with the conduction states. This assumption seems justified
for the systems under consideration whose 4$f$ valence deviates only
weakly from the integer value. We refer to this method as $f$-core
calculation. 
By using the $f$-core calculation for \YRS\ to interpret the results obtained on \LRS\ we rely on the facts that the lattice parameters agree within the experimental error\cite{Koehler2008} and that the results of the $f$-core calculation are independent of the 4$f$-occupancy.

The partially filled $f$ shell of the $4f^{13}$-configuration
necessarily carries a magnetic moment in agreement with Kramers' theorem.
The presence of local magnetic moments is reflected in the Curie-Weiss behavior observed at elevated temperatures in the magnetic susceptibility of \YRS\ and \YIS. The 4$f$ moments, however, interact only weakly with the conduction states
as can be inferred, \textit{e.g.}, from the low magnetic ordering temperature in \YRS.
We neglect the potential reconstruction of the conduction electron Fermi surface that may result from the long-range antiferromagnetic order and
account only for the 4$f$ charge which contributes to the potential seen by the conduction electrons. This amounts to effectively averaging over the local magnetic degrees of freedom in determining the self-consistent potentials. When comparing with experiment the bare bands derived from the effective potentials have to be renormalized by local 4$f$ excitations. Scattering off Crystalline Electric Field (CEF) excitations may enhance the effective masses and reduce the life-times of the conduction electrons. With these effects properly accounted for \cite{White1981,Fulde1983} the $f$-core model should quantitatively describe the electronic properties of Yb-based heavy-fermion compounds at elevated temperatures. However, for the Hall effect of the heavy-fermion compounds one has to take anomalous contributions into account which arise from the skew scattering at the local $f$-moments. These contributions may only be neglected at very low temperatures where, on the other hand, the $f$-core calculation is insufficient for the description of heavy-fermion compounds.
Rather, we shall use the Renormalized Band Calculation (RBC) to understand the Hall coefficient in the heavy-fermion compounds.

Treating the 4$f$ electrons as part of the ion core can be viewed as an extreme limit of an LDA+U calculation. Therefore, we shall compare our data with recent results obtained from LDA+U (Ref.~\onlinecite{Wigger2007}). The LDA+U calculation explicitly includes the magnetic moments of the 4$f^{13}$ configuration assuming long-range ferromagnetic order. This treatment preserves the translational invariance of the underlying lattice. It removes, however, the spin degeneracy of the conduction bands as they are split by the Zeeman effect. This splitting is rather small reflecting the weak coupling between the 4$f$ states and the conduction electrons. For this reason, we anticipate the energy bands of the LDA+U and the $f$-core calculation to agree in the low-energy regime, \textit{i.e.}, in the vicinity of the Fermi surface.

%
%
\subsubsection{Heavy Fermi liquid regime}

The strongly renormalized heavy quasiparticle bands are determined by means of the renormalized band method \cite{Zwicknagl1992,Fulde2006a} which combines material-specific \textit{ab initio} methods and phenomenological considerations in the spirit of Landau. The key idea is to construct an effective Hamiltonian for the low-energy excitations which uses the \textit{ab initio} potentials for the weakly correlated conduction electron channels while introducing one parameter to account for the specific local correlations among the 4$f$ electrons. The parameter is determined once by fitting to experiment and is kept fixed during subsequent investigations. A detailed description of the method and typical results for Ce-based compounds are given in Refs.\ \onlinecite{Zwicknagl1993a}, \onlinecite{Thalmeier2005}.
Operationally, it amounts to transforming the $f$-states of the spin-orbit ground state multiplet at the lanthanide site into the basis of CEF eigenstates $ \left| m \right\rangle  $  and introducing resonance-type phase shifts
 \begin{equation}
\tilde{\eta }_{f}(E)\simeq \arctan \frac{\tilde{\Delta}_{f}}{E-\tilde{\epsilon }_{f}}\label{eq:efren}
\end{equation}
where the resonance width $\tilde{\Delta}_{f}$ accounts for the renormalized
quasiparticle mass. The resonance energies $\tilde{\epsilon}_{fm}=\tilde{\epsilon}_f+\delta_{m}$ refer to the centers of gravity of the $f$-derived quasiparticle bands. Here $ \tilde{\epsilon}_f$ denotes the position of the band center corresponding to the CEF ground state while $\delta_m$ are the measured CEF excitation energies. One of the remaining two parameters,  $ \tilde{\epsilon}_f$, is determined by imposing the condition that the charge distribution is not altered significantly by introducing the renormalization. This makes the RBC a single-parameter scheme. The free parameter, $\tilde{\Delta}_{f}$, is adjusted in such a manner that the coefficient of the linear-in-$T$ specific heat at low temperatures is reproduced. The effective band structure Hamiltonian constructed along these lines corresponds to a hybridization model which closely parallels the one obtained from the periodic Anderson model in mean-field approximation. Alternatively, the RBC can be viewed as a parametrization scheme for the variation with energy of the real part of the local 4$f$ electron self-energy. The parameter to be determined by experiment is the slope at the Fermi energy while the value at $E_{\text F}$ is fixed by retaining the charge distribution.

The method has been shown to reproduce Fermi surfaces and anisotropies in the effective masses of a great variety of Ce-based compounds. In addition, it allows to predict Fermi liquid instabilities \cite{Pulst1993,Stockert2004,Thalmeier2005a,Zwicknagl2007,Eremin2008}.

In calculating the coherent 4$f$-derived quasiparticle bands in Yb-based heavy-fermion compounds we essentially follow the procedure for the Ce case as described above. We have to account for the fact that Yb can be considered as the hole analogue of Ce. Operationally this implies that we have to renormalize the 4$f$ $j=7/2$ channels at the Yb sites instead of the 4$f$ $j=5/2$ states in the Ce case. As the 4$f$ hole count is slightly less than unity the center of gravity $ \tilde{\epsilon}_f$ will lie below the Fermi energy. In addition, we have to reverse the hierarchy of the CEF scheme, \textit{i.e.},
\begin{equation}
\tilde{\epsilon}_f<0 \quad ;\quad \tilde{\epsilon}_{fm}=\tilde{\epsilon}_f-\delta_{m} \quad .
\label{eq:YbfRen}
\end{equation}

%
%
\subsection{Computational method}

The calculations are done on the basis of the experimental lattice parameters $a=b=\unit{4.007}\angstrom$, $c=\unit{9.858}\angstrom$ for \YRS\ and \LRS (cf.~section \ref{subsubsec:local_moment_regime}) and $a=b=\unit{4.032}\angstrom $, $c=\unit{9.826}\angstrom$
for \YIS\ (I-type).\cite{Trovarelli2000b,Hossain2005} 
The band structures were obtained by the fully relativistic formulation of the linear muffin-tin orbitals method \cite{Andersen1975,Skriver1984,Albers1986}. We adopt the atomic-sphere approximation including the combined correction term which contains the leading corrections\cite{Andersen1975}. In solving the band-structure problem, we include $s$-$p$-$d$-$f$-components at the Yb and the transition metal (Rh, Ir) sites and $s$-$p$-$d$-components at the Si sites. The spin-orbit interaction is fully taken into account by solving the Dirac equation. Although the relativistic effects hardly change the electron density distribution they nevertheless influence the actual location of the energy bands. This aspect is particularly important for the renormalized band structure since the spin-orbit splitting of the $d$-states is rather large on the energy scales relevant for the strongly renormalized heavy quasiparticles. Exchange and correlation effects were introduced using the Barth-Hedin potential \cite{Barth1972}. The band structure was converged for 405 \textbf{k}-points within the irreducible wedge, whose volume equals 1/16 of the Brillouin zone.
The density of states (DOS) was evaluated by the tetrahedron method
with linear interpolation for the energies. For the conduction band
the DOS was calculated at \unit{0.25}\milli\rydberg\ ($\approx \unit{3.4}\milli\electronvolt$) intervals. To obtain reliable values for the transport integrals
the energies were calculated at 2601 $\mathbf{k}$-points within the irreducible
wedge. Subsequently, the bands were interpolated using Mathematica
VI, and the result was used to numerically evaluate the desired quantities.

%
%
\subsection{Electronic structure}
\label{subsec:ES}

%
%
\subsubsection{\YRS\ in the local-moment regime}
\label{subsec:ES_YRS_fcore}
\begin{figure}
\includegraphics[height=\columnwidth, angle=270]{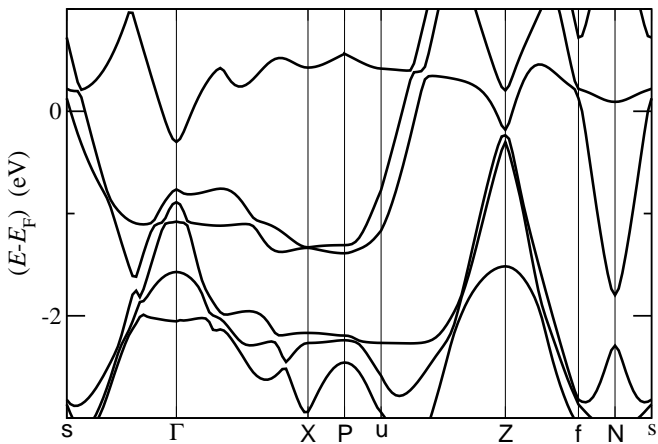}
\caption{YbRh$_2$Si$_2$: {Electronic} bands along symmetry lines with the Fermi energy $E_{\text F} =0$ chosen as reference energy. The Yb 4$f$ electrons are treated as part of the ion core. We follow the notation of Ref.~\onlinecite{Bradley1972} using the labels Z (0,0,1), $\Gamma$ (0,0,0), X (1,1,0), P (1,1,1) and N $\left( \frac{1}{2}, \frac{1}{2}, \frac{1}{2}\right) $ in units of $\left( \frac{\pi}{a},\frac{\pi}{a}, \frac{\pi}{a}\right) $. The labels s, f, and u refer to $( \tilde{a}, 0, 0) $ , $( \tilde{b}, 0, 2) $ and $( \tilde{b}, \tilde{b}, 2) $, respectively, with $\tilde{a}=1+( \frac{a}{c})^2$ and
$\tilde{b}=1-( \frac{a}{c})^2$.}
\label{fig:LDAfCoreBands}
\end{figure}
Figure \ref{fig:LDAfCoreBands} displays the {electron} bands of \YRS\ in the vicinity of the Fermi energy along symmetry lines with the 4$f$ electrons being treated as part of the ion core. Here, the band states have predominantly Rh 4$d$ character with some admixture of Yb 5$d$ character. The dispersion of \YRS\
agrees rather well with the results of recent LDA+U calculations \cite{Wigger2007,Jeong2006a}. In addition, it is consistent with energy bands deduced from photoemission studies.\cite{Wigger2007}

\begin{figure}
\includegraphics[height=\columnwidth, angle=270]{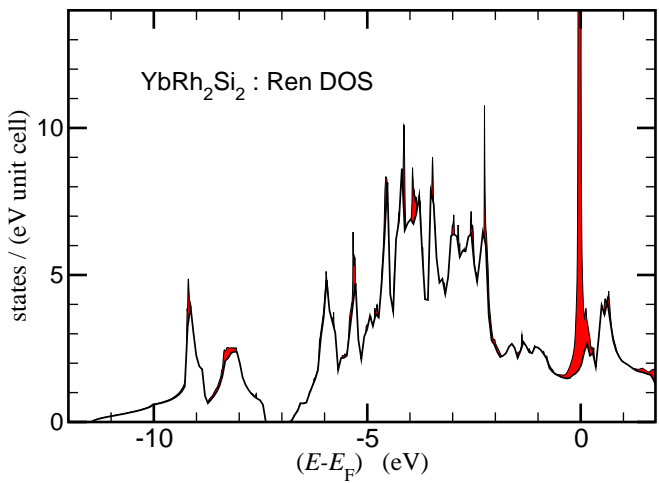}
\caption{(color online) \YRS: Comparison of the total DOS in the local moment regime ($f$-core calculation, solid line) and in the heavy-Fermi-liquid regime (RBC, red shaded area). The reference energy is the Fermi energy $E_{\text F}=0$. The two bands in the low-energy part are derived from the Si $s$ states. The dominant features are the Rh 4$d$ bands which hybridize with Si $p$ and Rh $s$ states near the bottom of the $d$ bands and with Si $d$ and Rh $p$ states near the top, respectively.}
\label{fig:YbRh2Si2fcoreVsRenDOS}
\end{figure}

In the 4$f$-core calculation, the broad bands intersecting the Fermi energy are exclusively formed by the non-$f$ conduction states. This is reflected in the low DOS at the Fermi energy $N(E_{\text F})=\unit{2.1} \usk \text{states} \per( \electronvolt \usk \text{unit cell})$ for the $f$-core calculation of \YRS\ as shown in Fig.\ \ref{fig:YbRh2Si2fcoreVsRenDOS}.

%
%
\subsubsection{\YRS\ and \YIS\ in the heavy-Fermi-liquid regime}
\label{subsubsec:electronic_structure_HF}
\begin{figure}
\includegraphics[width=\linewidth]{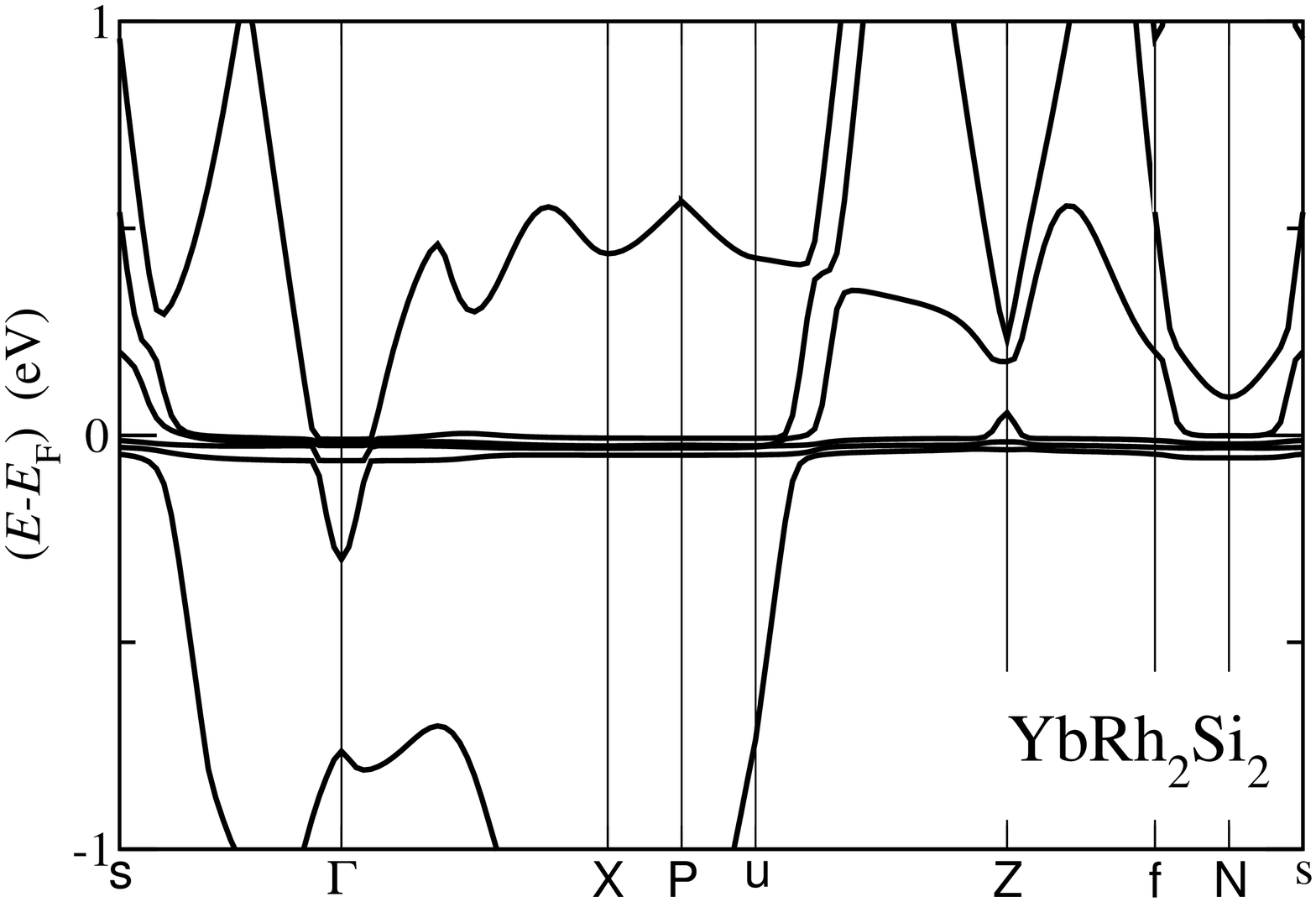}\\
\includegraphics[width=\linewidth]{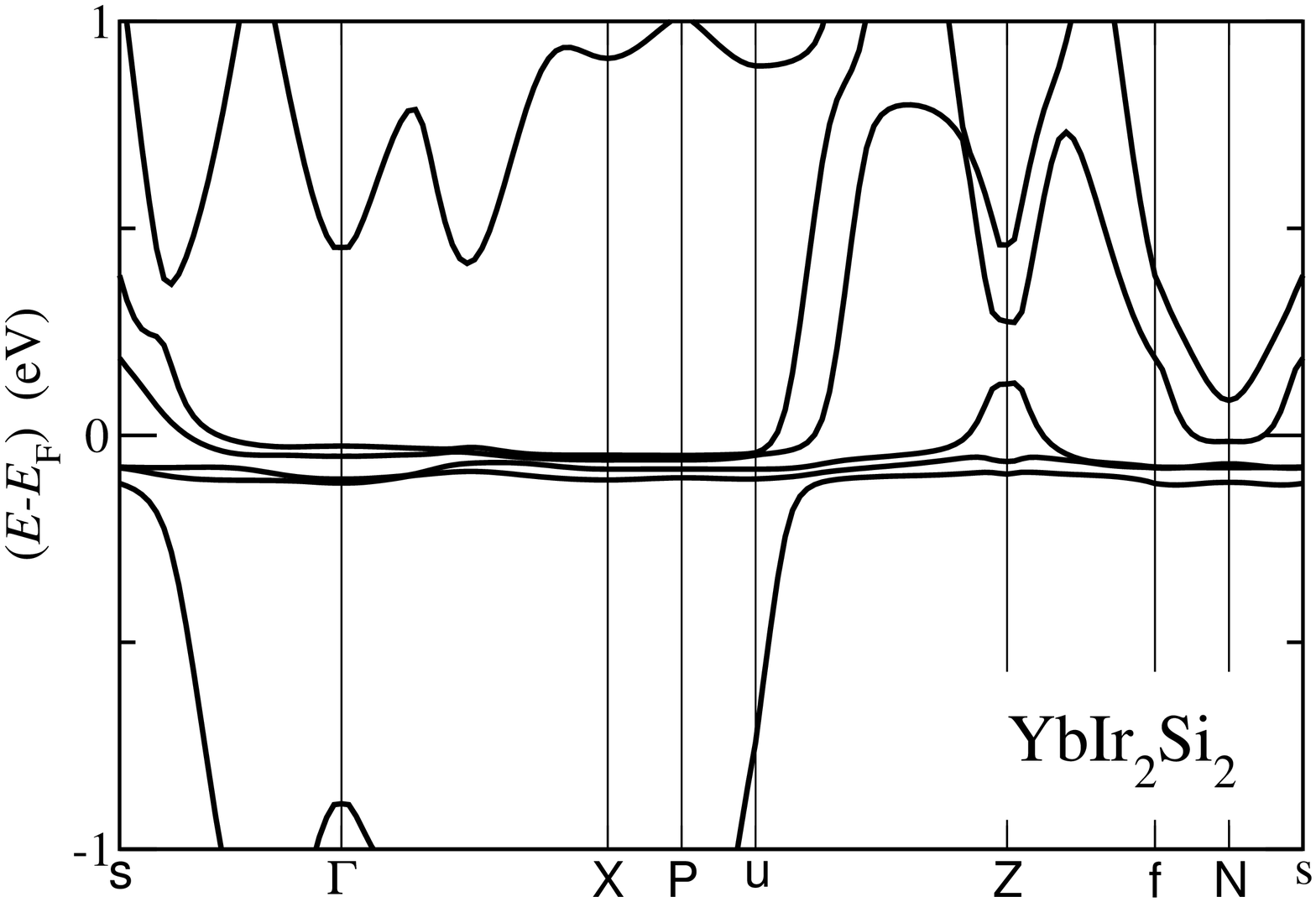}
\caption{Dispersion of the renormalized bands along symmetry lines for YbRh$_2$Si$_2$ (upper panel) and YbIr$_2$Si$_2$ (lower panel). The anisotropy of the CEF ground state leads to highly anisotropic hybridization strength which affects the relative shifts and the widths of the bands. The topology of the Fermi surfaces is mainly determined by the steep conduction bands. The symmetry of the CEF ground state  is reflected in effective mass anisotropies. The coordinates of the symmetry points Z, $\Gamma$, X, P, and N are specified in the caption of Fig.\  \ref{fig:LDAfCoreBands}}
\label{fig:RenBands}
\end{figure}

The calculations reported here adopt a CEF scheme which is consistent
with susceptibility and inelastic neutron-scattering data \cite{Hiess2006,Stockert2006}. 
The latter indicate that the 4$f$$^{13}$ states in
\YIS\ and \YRS\ are split
into 4 doublets with the energies \unit{0-18-25-36}\milli\electronvolt\ and \unit{0-17-25-43}\milli\electronvolt,
respectively. The low-energy properties are mainly determined by the
CEF ground state which is a superposition of $\left|j=7/2;j_{z}=\pm 5/2\right\rangle $ and $\left|j=7/2;j_{z}=\mp 3/2\right\rangle $ and which is well separated from the excited states. The CEF parameters and the CEF eigenstates are given in Ref. \onlinecite{Zwicknagl2009}. Using the effective quasiparticle {resonance} widths of $\tilde{\Delta}_f\simeq 20 \textrm{K}$ and  $\tilde{\Delta}_f\simeq 40 \textrm{K}$ as inferred from specific heat and thermopower measurements for the Rh (Ref.~\onlinecite{Gegenwart2006,Koehler2008}) and Ir (Ref.~\onlinecite{Hossain2005}) compounds yields the band structures displayed in Fig.\ \ref{fig:RenBands}. The dispersion of the renormalized bands of the Rh- and the Ir-compound are rather similar, the band widths scale with the characteristic temperatures. We shall concentrate on the results for the Rh-compound in the subsequent discussion.

\begin{figure}
\includegraphics[width=\columnwidth]{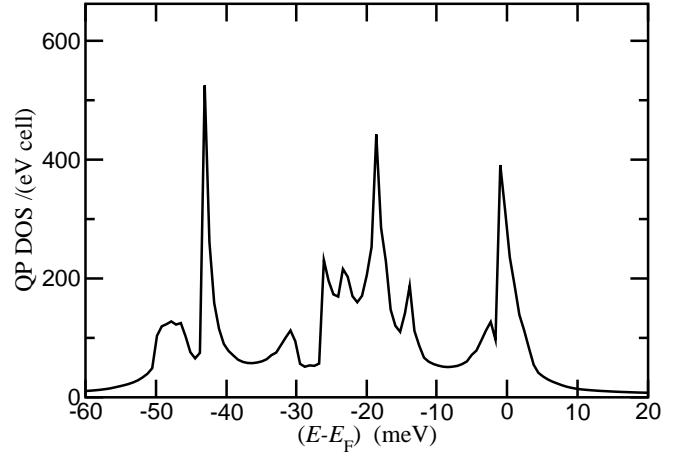}
\caption{Quasiparticle DOS for \YRS\ in the renormalized band structure calculation. For a comparison with the DOS plotted in Fig.~\ref{fig:LDAfCoreBands} consider the different scales.}
\label{fig:YbRh2Si2RenBandsLowEnergy}
\end{figure}

The RBC yields narrow $f$-derived quasiparticle bands in the vicinity of the Fermi energy, whereas the dispersion of the non-$f$ bands is essentially unaffected. This can be seen from Fig.\ \ref{fig:YbRh2Si2fcoreVsRenDOS} in which the DOS derived from the renormalized bands are compared with the $f$-core counterpart. The expanded view of the RBC DOS in the low-energy regime as depicted in Fig.\ \ref{fig:YbRh2Si2RenBandsLowEnergy} shows the contributions of the CEF-split 4$f$ states. The CEF excitations appear in the occupied part of the spectrum below the Fermi energy. The hybridization and hence the effective
quasiparticle masses are rather anisotropic. The renormalized band
calculations yield a DOS of $\unit{290} \usk \text{states} \per( \electronvolt \usk \text{unit cell})$ at $E_{\text F}$ corresponding
to specific heat coefficient $\unit{680} \milli\joule\usk\reciprocal\mole\usk\rpsquare\kelvin$. 
For \YIS\ a DOS of $\unit{48} \usk \text{states} \per ( \electronvolt \usk \text{unit cell})$ at $E_{\text F}$ is calculated corresponding to a Sommerfeld coefficient of  $\unit{113}\milli\joule\usk\reciprocal\mole\usk\rpsquare\kelvin$.

%

%
%
\subsection{Fermi surface and quasiparticles}

%
%
\subsubsection{Local-moment regime}
\begin{figure*}
	\begin{tabular}{rVV}
		&	$i=1$ `donut'	&	$i=2$ `jungle gym'	\\
		$f$-core 
		&
		\includegraphics[  width=0.75\columnwidth]{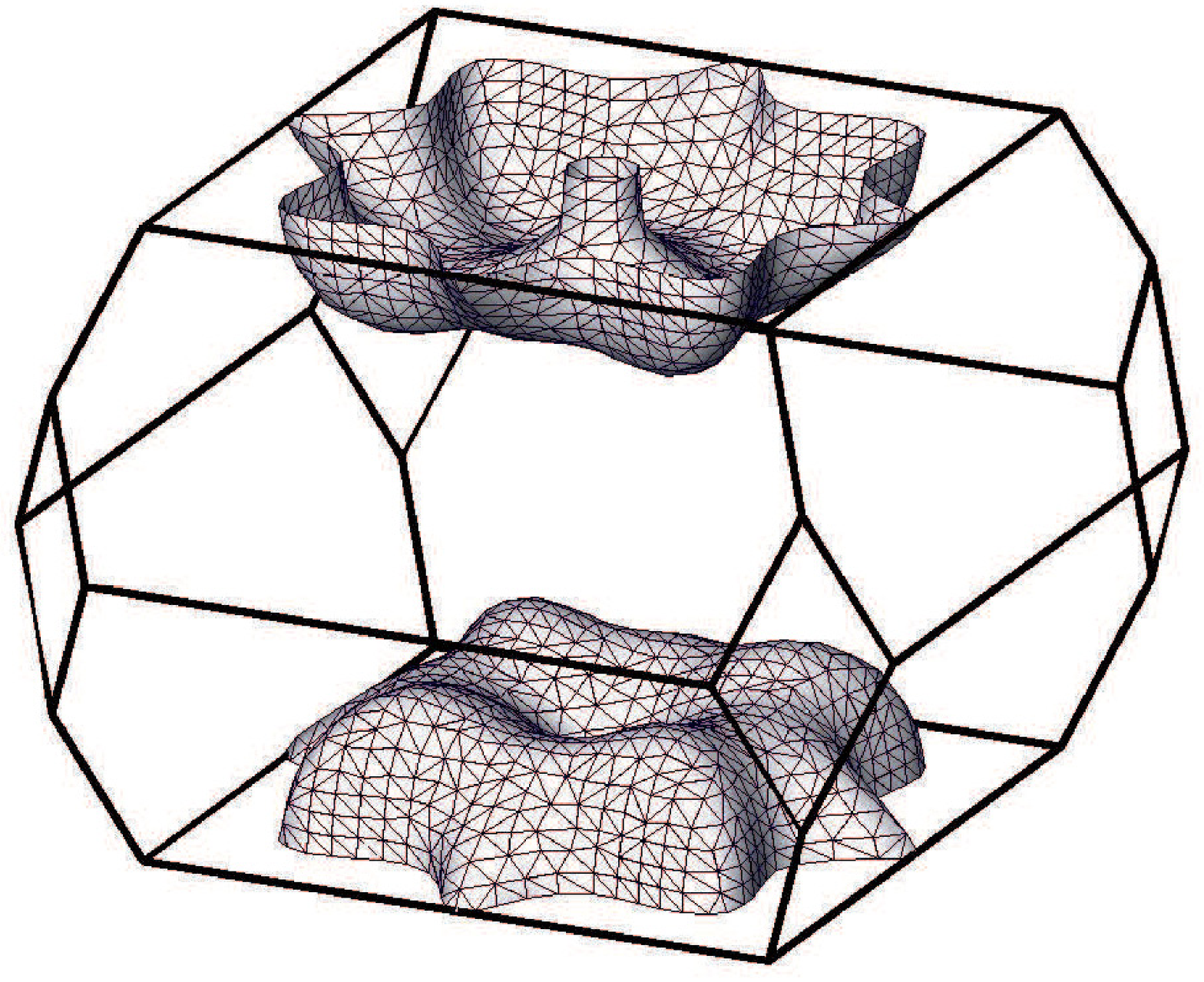}
		&
		\includegraphics[  width=0.75\columnwidth]{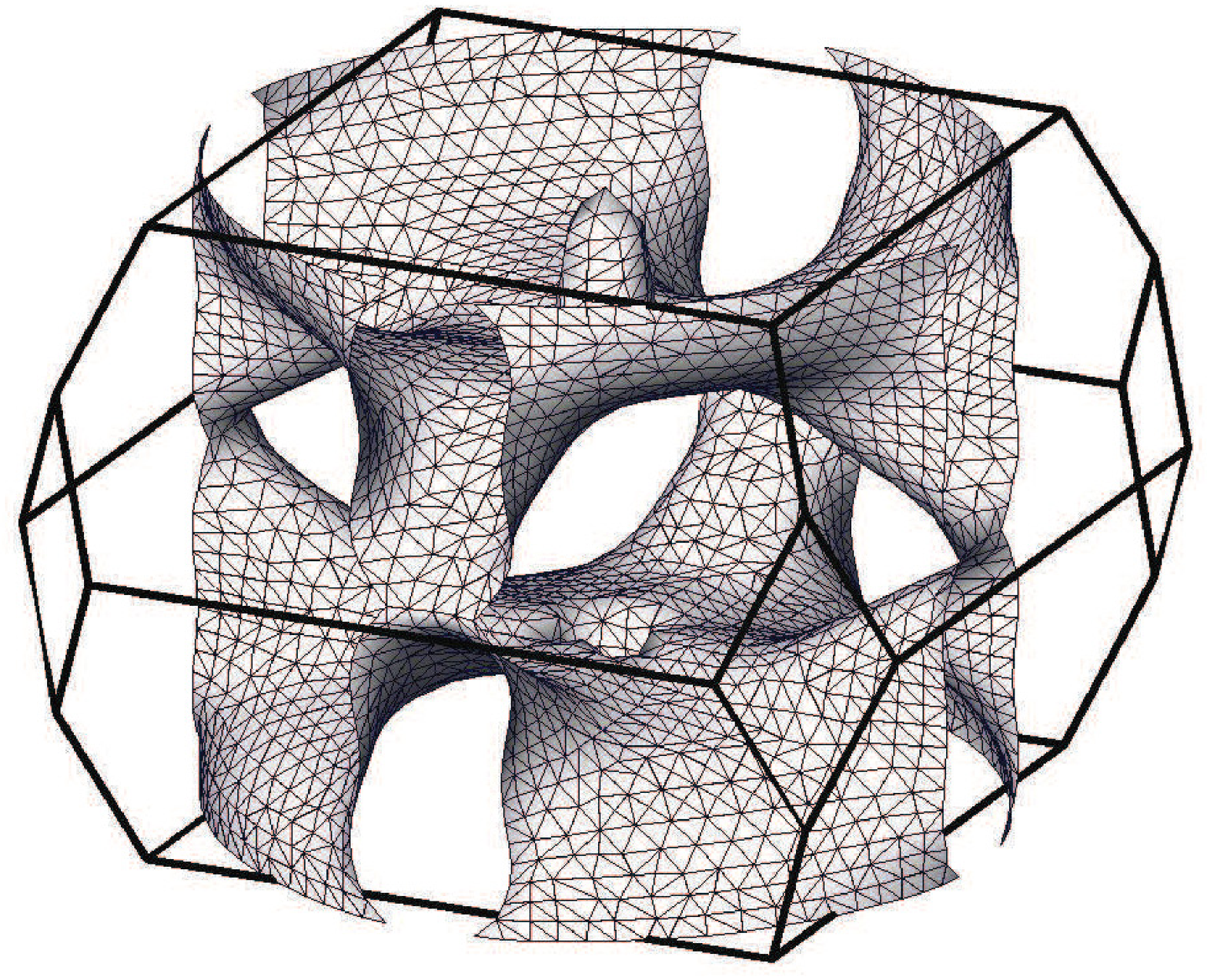}
		\\
		RBC
		&
		\includegraphics[  width=0.75\columnwidth]{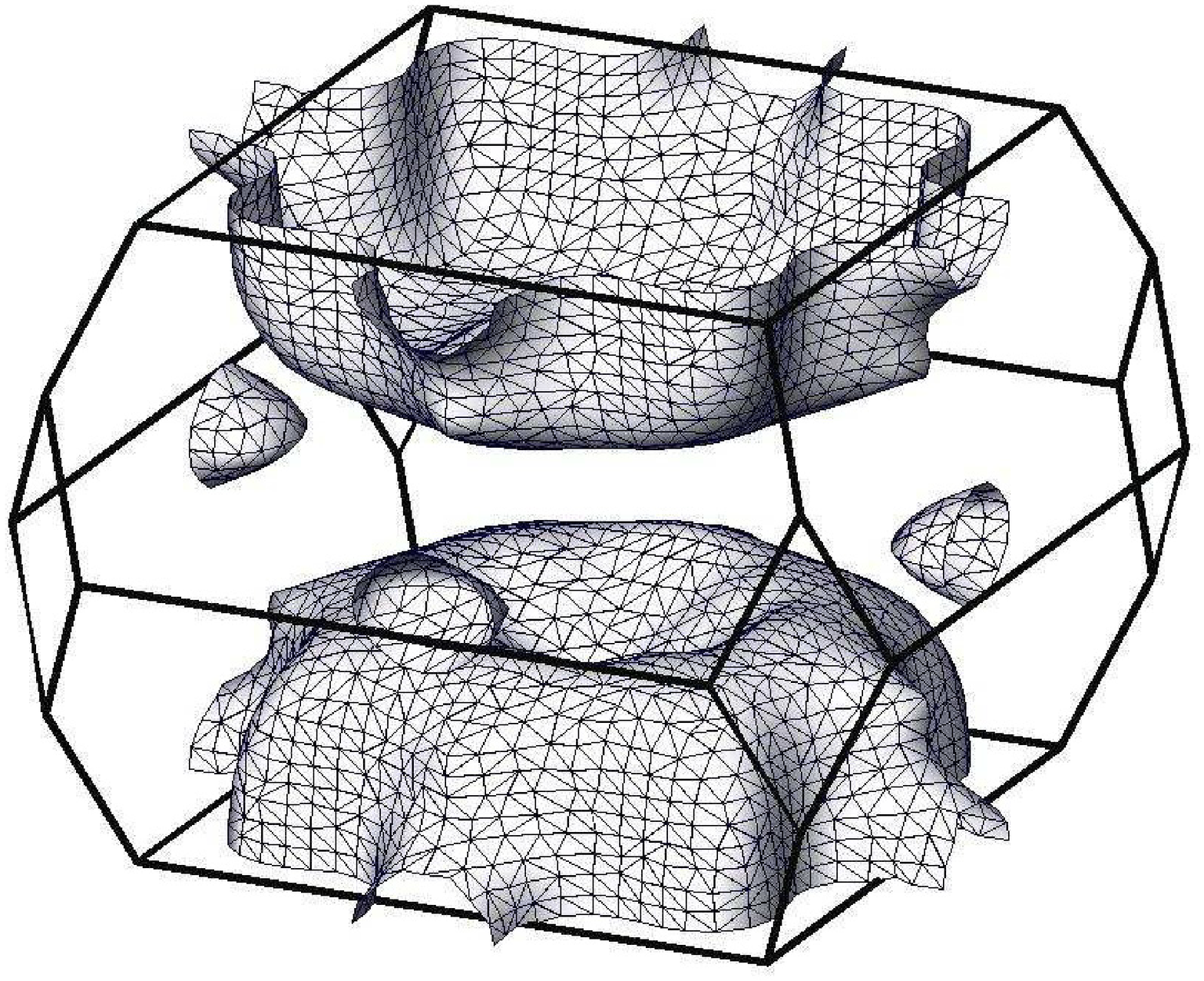}
		&
		\includegraphics[  width=0.75\columnwidth]{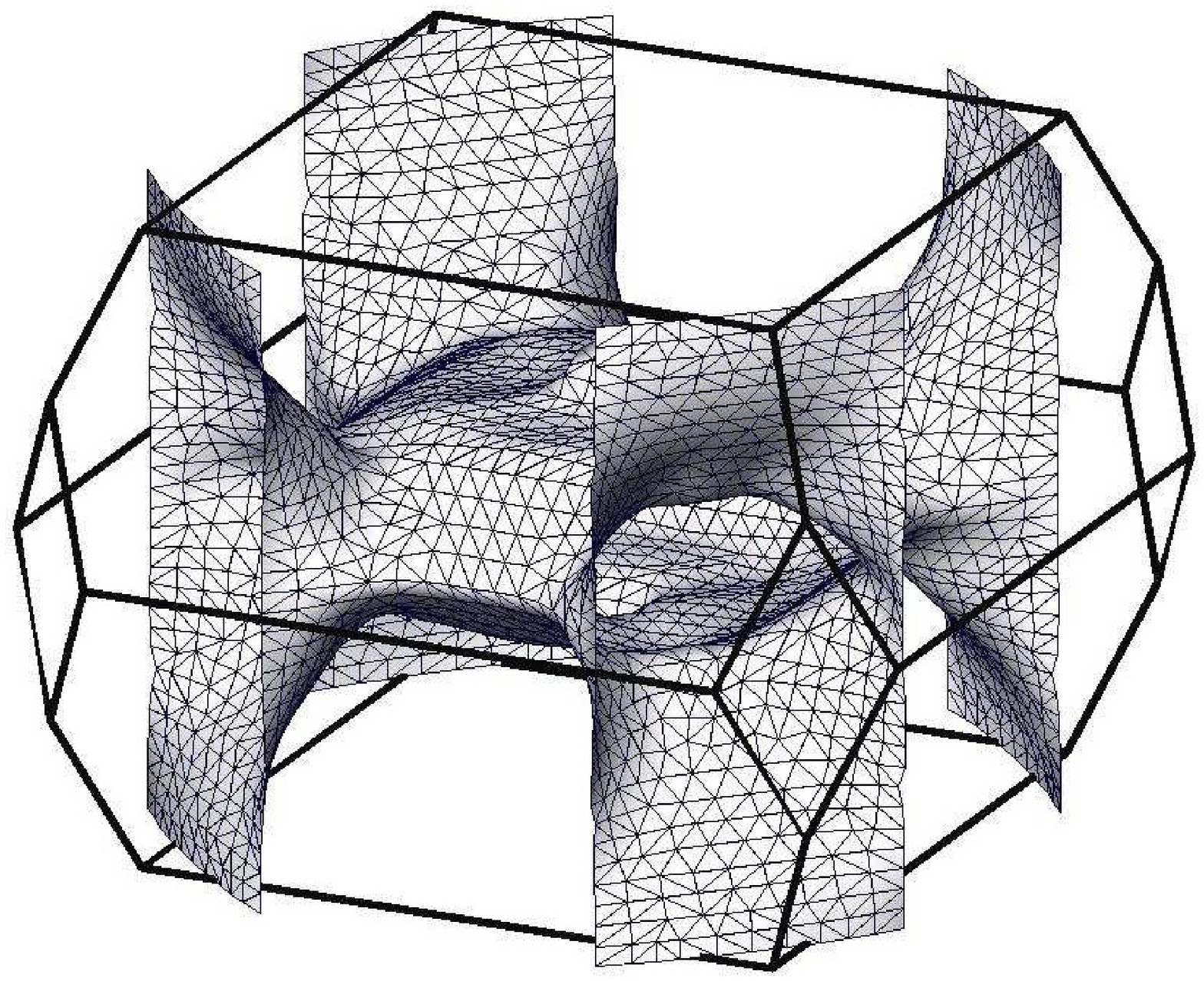}
		\\
	\end{tabular}
\caption{Calculated Fermi surfaces of \YRS: Major sheets of the Fermi surface in the $f$-core calculation representing the local-moment regime (top row) and the heavy-fermion regime (bottom row). In addition to the `donut' (left panel) and the `jungle gym' (right panel) there is a small electron surface (`pill box') which is not displayed here. The topology of the Fermi surface agrees well with previous findings.\cite{Wigger2007,Knebel2006,Jeong2006,Rourke2008}}
\label{fig:YbRh2Si2FSfCore}
\end{figure*}

The LDA calculation for localized 4$f$ electrons predicts three bands to cross the Fermi energy and leads to the Fermi surface which closely resembles previous results \cite{Wigger2007,Rourke2008}. It consists of three separate sheets. The two main sheets form a hole surface centered around the Z point, and a complex, multi-connected surface. Following Ref.~\onlinecite{Wigger2007} we shall refer to them as `donut' and `jungle gym', respectively. In addition, there is a small $\Gamma$-centered electron surface, the `pill box'. We shall focus on the two main sheets which are displayed in Fig.\ \ref{fig:YbRh2Si2FSfCore} as these two dominate the electronic properties. 
%

%
%
\subsubsection{Heavy-Fermi-liquid regime}

The RBC also predicts two major sheets at the Fermi
surface whose topologies resemble those found by LDA\cite{Jeong2006,Knebel2006}
(see Fig.\ \ref{fig:YbRh2Si2FSfCore}). The major sheets of the Rh-compound and its Ir-counterpart are rather similar. 
The main difference occurs in the small pockets: The small $\Gamma$-centered electron pocket of the Rh-compound is absent in the Ir-System where we find a Z-centered hole pocket instead.

From the comparison of the $f$-core results and those of the RBC it is obvious that they represent `small' and `large' Fermi surfaces, respectively. The difference of the Fermi volume accounts for the additional states related to the large quasiparticle DOS at $E_{\text F}$ for the heavy-Fermi-liquid limit (cf.\ Fig.~\ref{fig:YbRh2Si2fcoreVsRenDOS}). 

%

%
%
\subsection{Calculation of the Hall coefficient}
\label{subsec:Calc_HC}
For the chosen experimental geometry and using the Boltzmann approximation, the \HC\ is given in the low-field limit by \cite{Hurd1972}
\begin{equation}
\RH=\frac{\sum \limits_i \sigma _{xyz}(i)}{\left(\sum \limits_i \sigma _{xx}(i)\right)^2} \label{eq:HallCoefficientTetragonalSystem}
\end{equation}
where the conductivity tensor elements
\begin{equation}
\sigma _{xx}(i) =  e^{2}\frac{1}{\Omega }\sum _{\mathbf{k}}\tau (i)v_x^2(i,\mathbf{k}) \left(-\frac{\partial f}{\partial E(i,\mathbf{k})}\right)
\label{eq:SigmaNoH}
\end{equation}
and
\begin{multline}
\sigma _{xyz}(i)  =  \frac{e^{3}}{c}\, \frac{1}{\Omega }\sum _{\mathbf{k}}\tau ^{2}(i) \left[ v_x(i,\mathbf{k}) v_y(i,\mathbf{k}) \mathbf{M}_{yx}^{-1}(i,\mathbf{k}) \right. \\
\left. - v_x^2(i,\mathbf{k}) \mathbf{M}_{yy}^{-1}(i,\mathbf{k})  \right] \left(\fracpd{f}{E(i,\mathbf{k})}\right)
\label{eq:SigmaLinearInH}
\end{multline}
are summed over all bands $i$ intersecting the Fermi surface. We enumerate the `donut' and the `jungle gym' with $i=1$ and $i=2$, respectively. We restrict ourself to the first two bands which dominate the total conductivities and neglect small pockets. In the following, we allow for the possibility that the relaxation time $\tau$ may vary from one band to another but we shall neglect the variation of the relaxation time $\tau$ with wave vector $\mathbf{k}$ (with components $k_{\alpha}$). Here, $e$ and $c$ denote the electron charge and the vacuum speed of light, respectively. 
$\Omega $ represents the volume of the Brillouin zone and $f$ the Fermi distribution function. 
The components of the velocity
\begin{equation}
v_{\alpha }(i,\mathbf{k})=\frac{1}{\hbar }\frac{\partial }{\partial k_{\alpha }}E(i,\mathbf{k})
\label{eq:Velocity}
\end{equation}
and of the inverse mass tensor
\begin{equation}
\mathbf{M}_{\alpha \beta }^{-1}(i,\mathbf{k})=\frac{1}{\hbar ^{2}}\frac{\partial ^{2}}{\partial k_{\alpha }\, \partial k_{\beta }}E(i,\mathbf{k})
\label{eq:EffMassTensor}
\end{equation}
are deduced from the energy bands $E(i,\mathbf{k})$.

For the discussion of Fermi-surface effects we write the longitudinal and transverse conductivity as
\begin{align}
\sigma _{xx}(i) & = \sigma(i)\bar{\sigma }_{xx}(i) \nonumber \\
\sigma _{xyz}(i) & = \sigma_B(i)\bar{\sigma }_{xyz}(i) 
\label{eq:TransportIntegrals}
\end{align}
with the prefactors
\begin{align}
  \sigma(i)&=\frac{e^{2}}{m}\tau (i)\bar{n}(i)  \nonumber \\ 
  \sigma_B(i)& =\frac{\left|e\right|^{3}}{m^{2}c}\left(\tau (i)\right)^{2}\bar{n}(i) 
 \label{eq:Drude}
 \end{align}
being the Drude result for a gas of free particles with charge $\left| e\right|$. The particle density $\bar{n}(i)$ corresponds to the number of occupied states per unit cell in band $i$ while the reduced transport integrals $\bar{\sigma }_{xx}(i)$ and $\bar{\sigma }_{xyz}(i)$ account for the deviations of the conductivity tensor elements and the particle density from the free particle picture. The results for the systems under consideration are summarized in Table \ref{tab:CalculatedTransportIntegrals}.

\begin{table}[tb]
	\centering
	\caption{Calculated reduced transport integrals for the two different bands $(i=1,2)$. 
	The results derived for the two different Fermi-surface models are compared. 
	The Fermi surface results for \YIS\ within the RBC are included for comparison.
	See text for methods.
	\label{tab:CalculatedTransportIntegrals}}
		\begin{tabular}{llccccc}
			\toprule
			System& Method&$i$&$\bar{n}(i)$&$\bar{\sigma }_{xx}(i)$& $\bar{\sigma }_{xyz}(i)$ & $\bar{n}(i)\bar{\sigma }_{xyz}(i)$ \\
			\midrule
			\YRS &4$f$ core&1& 1.76 & 0.197 & $+0.289$ & $+0.50864$ \\
			& & 2 & 1.22 & 0.384 & $+0.153$ & $+0.18666$\\
			& & & & & & \\
			\YRS & 4$f$ RBC & 1& 1.37 & 0.0137 & $+0.00275$ & $+0.0037675$\\
			&& 2 & 0.63 & 0.0747 & $-0.00652$ & $-0.0041076$\\
			& & & & & & \\
			\YIS & 4$f$ RBC & 1 & 1.42 & 0.051 & $+0.00323$ & $+0.0045866$\\
			& & 2 & 0.58 & 0.138 & $-0.01003$ & $-0.0058174$ \\
			\bottomrule
		\end{tabular}
\end{table}

For the $f$-core calculations we obtain positive Hall (transverse) conductivity for both bands corresponding to hole-like character of the charge carriers. The crucial point is that for the RBC results, by contrast, the `jungle-gym', is predominantly electron-like as can be inferred from the reduced transport integrals listed in Tab.~\ref{tab:CalculatedTransportIntegrals}. 
Moreover, we find for \YRS\ that the two bands almost compensate each other. This is seen by the fact that the products $\bar{n}(i)\bar{\sigma }_{xyz}(i)$ of the two bands are close to each other in magnitude and of opposite sign. Their sum determines the numerator of Eq.\ \ref{eq:HallCoefficientTetragonalSystem}. Since we allow for different relaxation rates of the individual bands, this gives rise to a weighting of these two terms in the sum of Eq.\ \ref{eq:HallCoefficientTetragonalSystem}. Consequently, the total Hall coefficient very sensitively depends on the relative relaxation rates of the two bands. Even the sign of \RH\ may change if this balance is shifted only slightly toward the electron-like band. We shall discuss later that this might relate to the sample dependences observed in \YRS.

The calculated transport integrals vary only slightly with the position of the Fermi level.
This is contrary to the result of Ref.~\onlinecite{Norman2005} and reflects the different methods used: The LDA calculations of Ref.~\onlinecite{Norman2005} are not able to account for the position of the 4$f$ level with respect to the Fermi energy. The RBC on the other hand takes the correlation effects into consideration and, thus, does not rely on a shift of the 4$f$ level position.  

%
%
\section{Comparison to Hall effect measurements}

In this section we present the Hall-effect measurements and use the above results of the electronic structure calculations to advance our understanding of the experimental observations.
\label{sec:HE}

%
%
\subsection{Samples}
Single crystals of \YRS , \YIS\ and \LRS\ were synthesized applying an In flux-growth technique as described earlier.\cite{Trovarelli2000b} We note that within this work we concentrate on the I-type phase of \YIS\ which is isostructural to \YRS.\cite{Hossain2005}

In \LRS, also isostructural to \YRS, the Lu$^{3+}$ has 14 $f$ electrons and consequently retains a fully occupied $f$ shell without magnetic moment. Therefore, it serves as a non-magnetic reference compound to \YRS. An assignment of the \YRS\ $f$-core calculations to \LRS\ is justified by the fact that \LRS\ has equal lattice parameters within experimental error.
This allows us to model the experimentally observed temperature dependence of the Hall coefficient.
The $f$-core calculations yield a DOS (section \ref{subsec:ES_YRS_fcore}) which corresponds to a bare linear-in-$T$ specific heat coefficient of $\gamma \approx \unit{5}\milli\joule\usk\reciprocal\mole\usk\rpsquare\kelvin$ in good agreement with the experimental value $\gamma \approx \unit{6.5}\milli\joule\usk\reciprocal\mole\usk\rpsquare\kelvin$ found for \LRS\ (not shown). The resistivity as displayed in the inset of Fig.~\ref{fig:RHvsT_LRS} is approximately linear in $T$ above \unit{100}\kelvin\ with $\rho(\unit{300}\kelvin)=\unit{20}\micro\ohm\usk\centi\metre$. Both the specific heat and the resistivity indicate that \LRS\ is a simple non-magnetic intermetallic compound.  

\YRS\ and \YIS\ exhibit pronounced heavy-fermion behavior in various properties.\cite{Trovarelli2000b,Hossain2005}
In particular, the specific heat is largely enhanced (cf.\ section \ref{subsubsec:electronic_structure_HF}). However, their ground states differ: \YRS\ exhibits antiferromagnetic order at zero magnetic field, whereas \YIS\ is paramagnetic obeying Landau Fermi liquid (LFL) behavior below \unit{200}\mK.\cite{Hossain2005} Proximity of \YIS\ to a QCP is indicated by a logarithmic divergence of the specific heat for temperatures above \unit{200}\mK.
Since \YIS\ has a larger unit cell volume than \YRS\ it is assumed to be located on the paramagnetic side of the QCP as unit-cell expansion weakens magnetic ordering in Yb-systems. 
Consequently, \YIS\ serves as a reference compound with fully itinerant 4$f$-states as accounted for in the RBC. For \YRS, in its ground state in zero magnetic field, by contrast the $f$ electrons appear to be  localized as inferred from the Fermi surface reconstruction.\cite{Paschen2004}

%
%
\subsection{Experimental setup}

 All samples were polished to thin platelets of thickness in the range $\unit{25}\micro\metre \lesssim t \lesssim \unit{80}\micro\metre$. Subsequently, the samples were prescreened via resistivity $\rho(T,B)$ measurements to ensure In-free samples. 
The current $I$ was driven within the crystallographic $ab$ plane. The magnetic field $B$ was applied along the $c$ axis, thus, inducing the Hall voltage perpendicular to $I$ within the tetragonal plane, (see inset of Fig.\ \ref{fig:RHvsT_YIS}). To measure the Hall effect the transverse voltage $V_y$ was monitored. In order to cancel out magnetoresistance components due to contact misalignment, the Hall resistivity was obtained from the antisymmetric component of the field-reversed transverse voltage, $\rho_{\mathrm H}(B)=t\left[V_{\mathrm y}(+B)-V_{\mathrm y}(-B)\right]/2I$. The linear-response Hall coefficient $R_{\mathrm H}$ was derived as the slope of linear fits to the Hall resistivity $\rhoH$ for fields $B\leq \unit{0.4}\tesla$. Only the low-temperature Hall resistivity of \LRS\ displays a deviation from linearity as discussed in Ref.~\onlinecite{Friedemann2010}. In this case, the initial-slope Hall coefficient was deduced by extrapolating the differential Hall coefficient $\TRH(B)=\partial\rhoH/\partial B$ to ${B=0}$. Although this procedure yielded slightly larger values of \RH, the analysis presented here is not affected by this offset.

We note that the error on the absolute value of \RH\ arising from the uncertainty of the thickness of the samples is of the order of \unit{10}\percent. For \YRS, the results were scaled by a single factor in the temperature range $\unit{20}\kelvin \leq T \leq \unit{400}\kelvin$ to the previously published data \cite{Paschen2004}. The fact that this leads to a very precise match of \RT\ in this temperature range accounts for the error arising from the thickness which enters as a factor.  Consequently, the uncertainty of the sample thickness does not obstruct a detailed comparison of the different \YRS\ samples.

Measurements between \unit{2}\kelvin\ and \unit{400}\kelvin\ were conducted in a Quantum Design Physical Property Measurements System. For measurements down to $T=\unit{15}\mK$ a $^3$He/\,$^4$He-dilution refrigerator was utilized. In this case, the voltages were amplified by low-temperature transformers and subsequently recorded by a standard lock-in technique.

%
%
\subsection{Experimental Results and Discussion}
\subsubsection{\LRS}
\begin{figure}[tb]
 \includegraphics[width=\columnwidth]{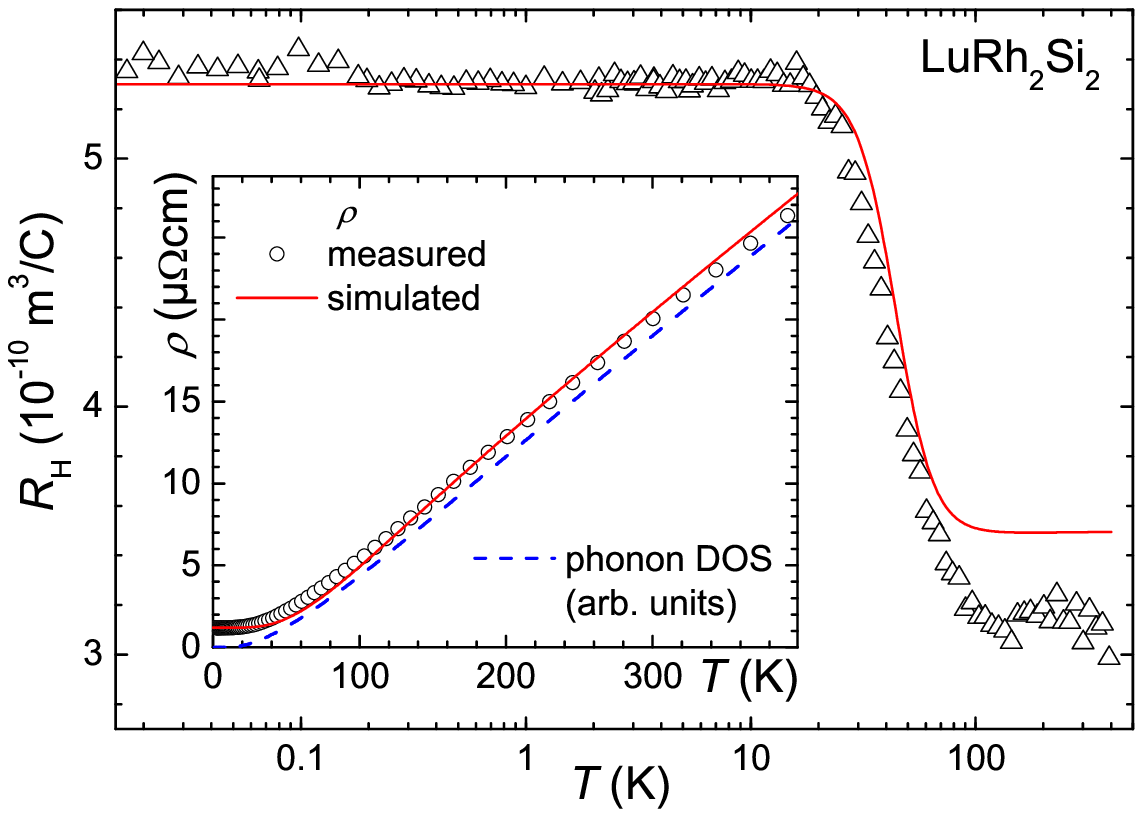}
	\caption{\label{fig:RHvsT_LRS}(color online) Temperature dependence of the linear-response Hall coefficient \RH\ of \LRS . The inset displays the resistivity of \LRS\ as a function of temperature. Solid lines represent the simulated data according to the two-band model described by Eqs.\ \ref{eq:TwoBandModelResistivity} to \ref{eq:GruneisenBloch}, using the parameters specified in Tab.\ \ref{tab:TwoBandModel_LRS} (see text). Dashed line in the inset denotes the electrical resistivity calculated on the basis of a measured phonon spectrum \cite{Stockert2006}. Adding the experimentally determined residual resistivity $\rho_{\text R}=\unit{1.2}\micro\ohm\usk\centi\metre$ in accordance with eq.~\ref{eq:Matthiessen} yields a precise match with the measured data.
	}
\end{figure}
The temperature dependence of the linear-response Hall coefficient, \RT, for \LRS\ is depicted in Fig.\ \ref{fig:RHvsT_LRS} for temperatures between \unit{17}\milli\kelvin\ and \unit{400}\kelvin.
For \LRS\ the Hall coefficient is always positive indicating hole-like charge carriers in agreement with the predictions of the $f$-core calculations on \YRS. The temperature dependence displays a constant value below \unit{20}\kelvin\ at $\RH \approx \HEunit{5.3}$ followed by a  crossover to another constant value of $\approx \HEunit{3.1}$ above \unit{100}\kelvin. We note that a very similar behavior of \RT\ has been observed for the non-magnetic La analogues of the Ce$M$In$_5$ ($M$=Co, Ir, Rh) family of compounds.\cite{Hundley2004}

For elemental copper a similar temperature dependence was observed. It was suggested to arise from two  bands effectively contributing to the Hall coefficient, with their weights changing as a function of temperature.\cite{Hurd1972} In the following we demonstrate that the combination of band-structure calculations and comprehensive electrical transport measurements allow to model the temperature dependence of \RH\ quantitatively.

\subsubsection{Application of a two-band model to \LRS}
In the light of the results of our calculation (Tab.~\ref{tab:CalculatedTransportIntegrals}) it is reasonable to interpret the temperature dependence of \RH\ of \LRS\ within a two-band model. Here, the observed crossover may be interpreted as the transition between the limits of the respective band dominating the total \HC. This may result from a shift of the relative scattering rate of the charge carriers in the individual bands off either phonons at high temperatures or static defects at low temperatures. For a quantitative analysis, we rewrite Eq.~\ref{eq:HallCoefficientTetragonalSystem} for two bands as a function of the resistivities as
\begin{equation}
	\RH \approx \rho_0^2 \sum_{i=1,2}{\frac{\RH(i)}{\rho(i)^2}}.
	\label{eq:TwoBandModelResistivity}
\end{equation}
Here, we approximated $\rho \approx \sigma^{-1}$ which is justified given the small Hall angle of less than \unit{3}\degree. The \HC\ of the individual bands is introduced as
\begin{equation}
\RH(i)=\frac{\sigma_{xyz}(i)}{\sigma_{xx}(i)^2}
\label{eq:RHi}
\end{equation}
The total resistivity $\rho_0$ given by
\begin{equation}
\rho_0^{-1} =\sum_{i=1,2}\rho(i)^{-1}
\label{eq:Res_Parallel}
\end{equation}
was measured simultaneously with the Hall effect and is displayed in the inset of Fig.\ \ref{fig:RHvsT_LRS}.
By introducing the ratio $r = \rho(1)/\rho(2)$ of the resistivities of the two bands we obtain the form 
\begin{equation}
	 \RH = \frac{\RH(1)+r^2 \RH(2)}{\left(r+1\right)^2}.
\label{eq:TwoBandModelRatio}
\end{equation}
 Here, it becomes obvious that the overall \HC\ is only a function of the ratio $r$ but not of the absolute values of $\rho(i)$, provided the $\RH(i)$ are temperature independent. For \LRS\ this latter assumption is supported by the band-structure calculations which yield constant values of $\RH(i)$ up to \unit{400}\kelvin, \textit{i.e.}, the thermal broadening of the Fermi surface has negligible influence, a typical behavior of conventional metals due to their high Fermi temperatures. We rather assume that merely $\rho(i)$ are temperature dependent. For \LRS\ we model the resistivity as a sum of different contributions according to Matthiessen's rule restricting ourself to a residual ($\rho_{\text R}$) and a phononic ($\rho_{\text P}$) term:
\begin{equation}
	\rho(i) = \rho_{\text R}(i)+\rho_{\text P}(i) 
	\label{eq:Matthiessen}
\end{equation}
The Bloch-Grüneisen law
\begin{equation}
	\rho_{\text P}(i)=C(i) \left(\frac{T}{\Theta_{\text D}}\right)^5\int\limits_0^{\Theta_{\text D}/T}{\frac{x^5}{\sinh^2(x)}\dd x}
	\label{eq:GruneisenBloch}
\end{equation}
describes the phononic component very well (cf. inset of Fig.~\ref{fig:RHvsT_LRS}).
This is corroborated by the agreement of the measured resistivity and the electrical resistivity calculated using a phonon DOS derived from measured inelastic neutron scattering spectra \cite{Stockert2006} (cf.\ Fig.~\ref{fig:RHvsT_LRS}).
In Eq.~\ref{eq:GruneisenBloch}, $C(i)$ is a constant related to the electron-phonon scattering probability of each band, and $\Theta_{\text D} = \unit{380}\kelvin$ is the Debye temperature determined from specific heat\cite{Ferstl2007}. 
Taking Eqs.~\ref{eq:TwoBandModelResistivity} to \ref{eq:GruneisenBloch} together one recognizes that the total Hall coefficient is determined at low temperatures by the ratio of the residual resistivities, and at high temperatures by that of the phonon scattering rates. This is in  good agreement with the experimental data: The low-temperature constant regime in \RT\ is observed in the temperature range where the resistivity is almost constant. By contrast, the high-temperature regime of \RT\ corresponds to a range where $\rho(T)$ appears to be dominated by electron-phonon scattering as indicated by the fact that $\rho(T)$ amounts to more than 10 times its residual value $\rho_{\text R}$. Finally, the crossover is centered at $T=\unit{50}\kelvin$ where the resistivity is twice its residual value implying that both contributions $\rho_{\text R}$ and $\rho_{\text P}$ are equal at this temperature.

Equations \ref{eq:TwoBandModelResistivity} to \ref{eq:GruneisenBloch} contain in total six free parameters: the Hall coefficients, the residual resistivities, and the phonon scattering rates of the two bands. In order to fit these equations to our data we proceeded as follows: Firstly, we utilized the results of our band structure calculation (Tab.~\ref{tab:CalculatedTransportIntegrals}) in Eq.~\ref{eq:RHi} to obtain the contributions $\RH(i)$ of the individual bands. These results are listed in the first column of Tab.~\ref{tab:TwoBandModel_LRS}. Secondly, these $\RH(i)$ and our experimental \RH\ are employed to obtain $r$ from Eq.~\ref{eq:TwoBandModelRatio}. This step is performed with the value of \RH\ measured at low temperatures yielding $r=1.3$, as well as in the high-temperature limit. However, in the latter case no 
exact solution is possible since the solution space is limited to $\RH \geq \HEunit{3.5}$ where $r=5.7$ for the values $\RH(i)$ obtained on the basis of the calculated electronic structure. The discrepancy between the measured and the calculated \HC\ at high temperatures may also be corrected by a change of $\RH(2)$ to $\HEunit{3.6}$. This might indicate that the assumption of an isotropic relaxation time is not fully justified. However, we rather stick to the results of the band-structure calculation as any change would be arbitrary. 
Thirdly, we take advantage of the fact that, at our lowest measurement temperature ($T \ll \Theta_{\text D}$), $\rho_{\text P}$ is negligible leaving only $\rho_{\text R}$ in Eq.~\ref{eq:Matthiessen}. With $r$ known from the second step and the total resistivity (Eq.~\ref{eq:Res_Parallel}) set to the experimentally obtained value at low temperature, the individual $\rho_{\text R}(i)$ can be calculated. In the high-temperature regime, on the other hand, the residual term in Eq.~\ref{eq:Matthiessen} is negligible and hence, the individual values of $\rho_{\text P}(i)$ are obtained from which, in turn, $C(i)$ is inferred. All results are summarized in 
Tab.\ \ref{tab:TwoBandModel_LRS}.

\begin{table}
	\centering
	\caption{Parameters calculated for \LRS\ within the two-band model. The values were obtained from the parameters of Tab.~\ref{tab:CalculatedTransportIntegrals} following the recipe outlined in the text.
	Inserting the values listed here, we simulate \RT\ and $\rho(T)$ as shown by the solid lines in Fig.~\ref{fig:RHvsT_LRS}.}	
		\begin{tabular}{c c c c}
		\toprule
		$i$ 	& $\RH(i)$ & $\rho_{\text R}(i)$ & $C(i)$ \\
		 &	$(\unit{10^{-10}}\cubicmetre\per\coulomb)$ & (\micro\ohm\usk\centi\metre) & (\micro\ohm\usk\centi\metre) \\
		\midrule
			1		&	21	& 2.75 & 7 \\
			2		& 4.2 & 2.11 & 1.23 \\
		\hline
		\multicolumn{2}{c}{resistivity ratio $r$} & 1.3 & 5.7 \\
		\bottomrule
		\end{tabular}
	\label{tab:TwoBandModel_LRS}
\end{table}

With the parameters of Tab.\ \ref{tab:TwoBandModel_LRS} we are now in the position to simulate the overall temperature dependence of both the Hall coefficient and the resistivity again employing Eqs.\ \ref{eq:TwoBandModelResistivity} to \ref{eq:GruneisenBloch}. The results are included in Fig.~\ref{fig:RHvsT_LRS} as solid lines. 
The good quantitative and the even better qualitative agreement of the simulated and measured data justify the application of the two-band model. We wish to emphasize that the position of the crossover in \RT\ and the position where $\rho(T)$ deviates from its residual value are not fitted but are dictated by the Debye temperature which was determined independently.

\subsubsection{\YIS\ and \YRS}
\begin{figure}[tb]
  \centering
  \setlength{\unitlength}{\linewidth}
	\begin{picture}(1,.715)
  		\put(0,0){\includegraphics[width=\unitlength]{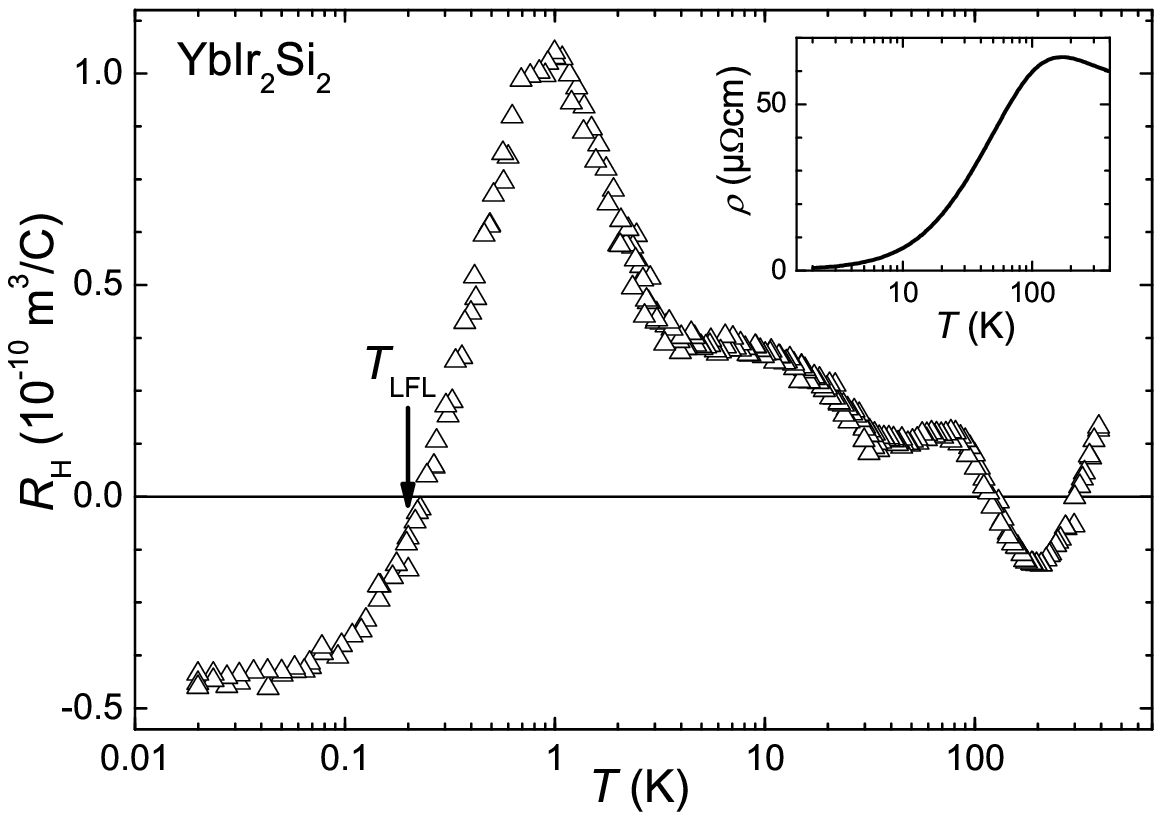}}
  		\put(.35,.1){\includegraphics[width=.5\unitlength]{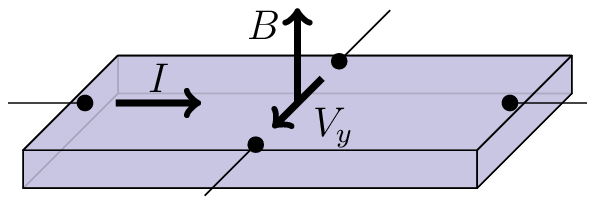}}
  \end{picture}	
	\caption{\label{fig:RHvsT_YIS}(color online) Initial-slope Hall coefficient \RH\ of \YIS. The arrow marks the temperature below which LFL behavior was observed\cite{Hossain2005}. Upper inset displays the temperature dependence of the resistivity. Lower inset sketches the setup.}
\end{figure}

\begin{figure}
	\centering
	{\includegraphics[width=\linewidth]{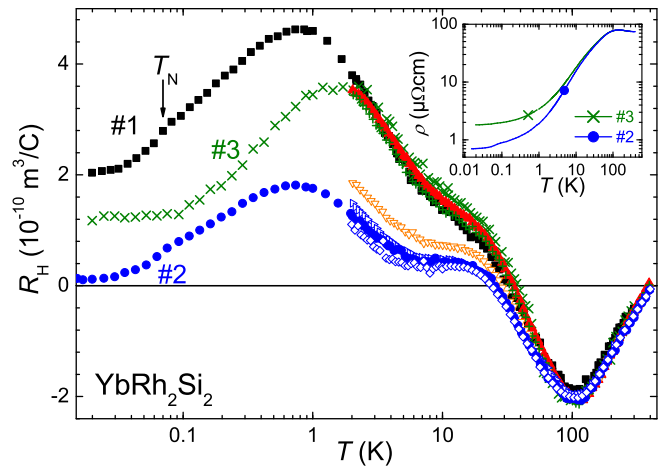}}
	\caption{(color online) Temperature dependence of the \HC\ for different samples. See Tab.~\ref{tab:RHvsT_YRS} for legend. Results for samples of the same batch are shown in identical color. Arrow indicates the Néel-temperature. Inset displays the resistivity of two selected crystals.}
	\label{fig:RHvsT_YRS}
\end{figure}

For the heavy-fermion compounds \YIS\ and \YRS\ the temperature dependence of the Hall coefficient is more complicated as can be seen from Figs.\ \ref{fig:RHvsT_YIS} and \ref{fig:RHvsT_YRS}. At high temperatures both compounds show a minimum in \RT\ in the same temperature range where the resistivity assumes a maximum (cf.\ insets of Figs.~\ref{fig:RHvsT_YIS} and \ref{fig:RHvsT_YRS}), namely at approximately \unit{180}\kelvin\ for \YIS\ and \unit{120}\kelvin\ for \YRS. 
This corroborates the earlier assignment of this minimum in \RT\ of \YRS\ to the anomalous \HE\ arising from skew scattering which predicts such a correlation between the resistivity and the anomalous Hall contribution.\cite{Paschen2005}

Between \unit{80}\kelvin\ and \unit{30}\kelvin\ \RT\ of \YIS\ assumes a plateau at a value of \HEunit{0.14}. This indicates that the anomalous contribution, typically being of importance around the resistivity maximum only, is superposed to a normal component as expected in the theory of the anomalous Hall effect.\cite{Fert1987}

In the temperature range between \unit{30}\kelvin\ and \unit{8}\kelvin\ a crossover to another plateau at a value of \HEunit{0.35} is observed in \RT\ of \YIS.
Two possible reasons may account for this observation: (i) The crossover might be of the same two-band nature as in \LRS. However, as the crossover in \YIS\ is situated at a lower temperatures it should be accompanied by a decreased value of the Debye-temperature. Unfortunately, $\Theta_{\text D}$ is not yet known. Moreover, single crystals of LuIr$_2$Si$_2$ are not available to look for a possible shift of the two-band crossover in this non-magnetic reference compound.
(ii) Alternatively, the crossover might manifest the Fermi surface change arising from the onset of the Kondo screening effect which leads to itinerant $f$ electrons contributing to the Fermi surface at low temperatures.

Below \unit{4}\kelvin\ the Hall coefficient of \YIS\ exhibits a pronounced increase, peaks at \unit{1}\kelvin\ and drops at lower temperatures. At \unit{0.23}\kelvin, \RT\ changes sign and finally saturates at the lowest temperatures at a value of \HEunit{-0.4}. 

For \YRS\ the minimum in \RT\ at \unit{100}\kelvin\ caused by the anomalous \HE\ is uniquely observed for all samples investigated. By contrast, below \unit{50}\kelvin\ strong sample dependences are present. Figure \ref{fig:RHvsT_YRS} consists of data obtained for a large variety of samples. Three of theses samples were selected for low-temperature measurements down to \unit{15}\mK\ and represent the full range of sample dependences. Sample 1 and 3 exhibit a shoulder in \RT\ around \unit{15}\kelvin\ whereas \RT\ of sample 2 shows a plateau in the temperature interval $\unit{7}\kelvin \leq T \leq \unit{20}\kelvin$.
All samples exhibit a maximum in \RT\ around \unit{1}\kelvin\ like in \YIS, however, at different absolute values.
This maximum is assigned to the quantum critical spin fluctuations operating for all samples in the same temperature regime.
Such a pronounced extremum was also reported for Ce$M$In$_5$ where antiferromagnetic fluctuations were suggested as the microscopic origin.\cite{Hundley2004,Nakajima2006} In \YRS , NMR investigations \cite{Ishida2002} revealed antiferromagnetic fluctuations to be present in the designated temperature range. However, the temperature dependence of $\RT \propto T^{-1}$ predicted in Ref.~\onlinecite{Nakajima2006} is not observed in \YRS\ nor \YIS .

Finally, at the lowest temperatures all samples of \YRS\ show a saturation of \RT, setting in just below the Néel temperature. However, the saturation value appears to be sample dependent.


\begin{table}
	\begin{tabular}[b]{c c l c c}
			\toprule
			 &\mbox{sample}  & batch & $\frac{\rho(\unit{300}\kelvin)}{\rho(\unit{2}\kelvin)}$& $\RH(\unit{2}\kelvin)$ \\
			\midrule
			 \includegraphics[height=\SymbolHeight]{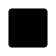} 				& \#1	& 37105 	& 20.7 	& 3.8	\\
			 \addlinespace[.75em]
			 \includegraphics[height=\SymbolHeight]{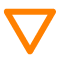}	 		& 		& 63111 	& 20.7 	& 1.9	\\
			 \addlinespace[.75em]
			 \includegraphics[height=\SymbolHeight]{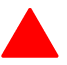} 				& 		& 63113 	& 15.3 	& 3.5	\\
			 \addlinespace[.75em]
			 \includegraphics[height=\SymbolHeight]{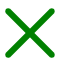} 						& \#3	& 63114 	& 15.2	& 3.4	\\
			 \includegraphics[height=\SymbolHeight]{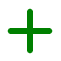} 						& 		& 63114 	& 16.1 	& 3.7	\\
			 \addlinespace[.75em]
			 \includegraphics[height=\SymbolHeight]{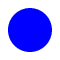}						& \#2	& 63116 	& 24.0	& 1.3	\\
			 \includegraphics[height=\SymbolHeight]{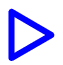}		& 		& 63116 	& 24.2	& 1.5	\\
			 \includegraphics[height=\SymbolHeight]{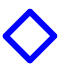}				& 		& 63116 	& 26.5	& 1.1	\\
			\bottomrule
		\end{tabular}
	 \caption{Sample and batch numbers of the data sets shown in Fig.~\ref{fig:RHvsT_YRS} together with the value of the \HC\ (in units of \unit{10^{-10}}\metre\cubed\per\coulomb) and the resistivity ratio at \unit{2}\kelvin. }%
	\label{tab:RHvsT_YRS}
\end{table}

\subsubsection{Sample Dependences}

A series of \YRS\ samples with different residual resistivities have been investigated above \unit{2}\kelvin. It turns out that the saturation values at lowest temperature correlate with the values at the plateau/shoulder around \unit{20}\kelvin. The plateau is more pronounced for samples with a lower saturation value:
Sample 1 obeys a comparably slight shoulder and saturates at the highest low-temperature value. 
Sample 2 depicts the most pronounced shoulder resembling a plateau and exhibits the lowest saturation value.
Sample 3 with an intermediate saturation value obeys a more pronounced shoulder than sample 1. 
The correlation indicates that the maximum around \unit{1}\kelvin\ is caused by a superposed contribution which itself is not affected by the sample dependences.

It is observed that samples from the same batch (cf. colors in Fig.~8) show almost identical \RT\ curves. From this we infer that sample dependences arise from slight differences in crystal growth.
By contrast, a correlation between sample quality and these sample dependences can not be found. This is quantitatively analyzed in Tab.\ \ref{tab:HallCoefficients} for the selected samples by a comparison of the low-temperature saturation value of \RH\ with the residual resistivity ratio ($RRR$). An equivalent conclusion is found for the larger set of samples: In the absence of measurements in the \mK-range, the resistivity ratio at \unit{2}\kelvin\ is used to quantify the sample quality which appears to be uncorrelated with the Hall coefficient at \unit{2}\kelvin.

\begin{table}[tb]
\center
\caption{Calculated and experimental \HC s. The calculated \RH\ values are derived by considering the two major bands with the assumption of equal relaxation times (see text). 
Values of the zero-temperature \HC , \RHO , extrapolated from measurements are given for the related materials. \RH\ is in units of $(\unit{10^{-10}}\cubicmetre\per\coulomb)$. \label{tab:HallCoefficients}}
\begin{tabular}{llc|ccc}
\toprule
\multicolumn{3}{c|}{Calculation} & \multicolumn{3}{c}{Experiment} \\
System & Method & \RH & Sample &$RRR$ & \RHO \\
\midrule
\YRS & 4$f$ core & 5.16 & \LRS & 17 & 5.3\\
& & & & \\
& & & \YRS & \\
\YRS & 4$f$ RBC & -0.39 & sample 1 & 70 & 2.0 \\
& & & sample 2 & 120 & 0.1   \\
& & & sample 3 & 40 & 1.2   \\
& & & & \\
\YIS & 4$f$ RBC & -0.26 & \YIS\ & 325 & -0.4 \\
\bottomrule
\end{tabular}
\end{table}

No sample dependences were observed for \LRS\ for which three samples where investigated. In the case of \YIS\ only one sample without indium enclosures could be identified. In samples with indium enclosures the rearrangement of the current distribution largely disturbs the Hall-effect measurement and therefore no statement on sample dependences can be made. However, the fact that the calculated and the measured \HC\ agree suggests that the measurements depict the intrinsic behavior.

\subsubsection{Comparison of Theory and Experiment}
Unfortunately, for \YIS\ and \YRS\ it is not possible to apply the two-band model as done for \LRS\ because both the resistivity and the Hall coefficient contain additional (quantum critical) contributions. Thus, the number of unknown parameters would increase and could in particular not be mapped with measured quantities. For a qualitative discussion we make the simplified assumption of equal relaxation rates for the two bands which yield the Hall coefficients listed in Tab.\ \ref{tab:HallCoefficients}.


For \YIS, the agreement between the zero-temperature \HC , \RHO , extrapolated from measurements, and the calculated value is remarkable (see Tab. \ref{tab:HallCoefficients}). 
{In the case of \YRS\ our band structure calculations predict a value lower than the experimental \RHO. This might be due to deviations from equal relaxation rates as the sample dependences indicate that small changes can have large influence.} 

The most straight-forward interpretation of the sample dependences in \YRS\ arises from the insight provided by the band-structure calculations. As shown in section \ref{subsec:Calc_HC}, the two bands dominating the Hall coefficient are of opposite character and almost compensate each other. The actual value of the total Hall coefficient, therefore, depends sensitively on the ratio of the scattering rates of the individual bands because they enter as a weighting factor in the summation of the individual contributions. Hence, it is reasonable to assign the observed sample dependences to changes of the relative scattering rates. This is in agreement with the fact that other properties like specific heat, susceptibility and even resistivity (cf.~inset of Fig.~\ref{fig:RHvsT_YRS}) do not obey such strong sample dependences as none of these properties depends this sensitively on the ratio of the scattering rates. In fact, the resistivity is a sum of the two (Eq.~\ref{eq:Res_Parallel}).

Finally, the fact that samples of the same batch exhibit almost identical behavior in \RT\ allows us to surmise that the sample dependences are related to tiny differences in the actual stoichiometry caused by different crystal growth conditions. Such sensitivity on minute changes of the composition is known, for instance, for the heavy-fermion superconductor CeCu$_2$Si$_2$ where it leads to even more dramatic effects, which include drastic changes in the ground state.\cite{Steglich2001a}

{The sample dependences in \RT\ are observed to set in around \unit{70}\kelvin. Below \unit{10}\kelvin\ they are fully developed and appear to be conserved down to the lowest temperatures as an offset between different samples. Consequently, the low-temperature Hall coefficient reflects the Fermi surface with sample-dependent, but fixed, weight of the individual sheets. This indicates that the Hall crossover, monitoring the Fermi surface reconstruction at the QCP, is robust against sample dependences as indeed observed \cite{Friedemann2009a}.}

{The comparison of the calculated Hall coefficient for the limiting cases of localized (4$f$ core) and itinerant (4$f$ RBC) 4$f$ electrons in Tab.~\ref{tab:HallCoefficients} shows that the inclusion of the Yb 4$f$ states into the Fermi volume leads to a decrease of the Hall coefficient. Consequently, the finding of a jump from larger \RH\ at zero field towards a lower value at elevated fields in isothermal scans\cite{Paschen2004,Friedemann2009a} indicates a localization of the $f$ electrons on the low field side of the QCP in \YRS.}

%
%
\section{Conclusion}
We have calculated the electronic band structure of \YRS\ and \YIS\ both with and without taking the Kondo scattering into account. Two bands were found to dominate the Hall coefficient. Both these bands are hole-like in the case of the $f$-core calculations neglecting the Kondo effect but are of opposite character for the case of the renormalized band calculation.
The derived results allow for an in-depth analysis of the Hall coefficient of the non-magnetic reference compound \LRS. We are able to quantitatively understand the temperature dependence of the Hall coefficient in terms of a two-band model.

Furthermore, we present Hall effect measurement on \YIS. Here, the temperature dependence of the Hall coefficient parallels many features known for \YRS. In particular, the anomalous contribution is seen to follow the expected trend. Remarkably, the Hall coefficient derived from the renormalized band calculation is in very good agreement with the measured value at lowest temperatures.

Finally, the sample dependences of the low-temperature Hall coefficient of \YRS\ are discussed in terms of the two bands predicted by the calculations and seen in \LRS.
The fact that the renormalized band calculation predicts the two bands to almost compensate each other indicates that the sample dependences arise from small changes of the scattering rates for the individual bands. These changes are ascribed to minute differences in the sample composition as samples of the same batch show almost identical behavior.
More importantly, {despite the strong sample dependencies our comprehensive study} on \YRS\ {confirms that }the distinct change of the Hall coefficient, found in isothermal scans across the quantum critical point {marks} a substantial change of the Fermi surface {in the same} way as expected in the Kondo breakdown scenario.\cite{Friedemann2009a}

%
%
\begin{acknowledgments}
The authors would like to thank P. Gegenwart, S. Kirchner, Q. Si, R. Valent{\'i}, T. Westerkamp and G. Wigger for fruitful discussions. This work was partially supported by NSF-DMR-0710492, NSF-PHY-0551164, FP7-ERC-227378 and DFG Forschergruppe 960.
\end{acknowledgments}


\end{document}